\documentclass[superscriptaddress,preprint,amsmath,amssymb,aps,prb]{revtex4-2}

\usepackage{graphicx}
\usepackage{dcolumn}
\usepackage{bm}
\usepackage{xcolor}
\usepackage[mathlines]{lineno}

\let\oldsubequations\subequations
\let\oldendsubequations\endsubequations

\renewenvironment{subequations}
  {\linenomathNonumbers\oldsubequations}
  {\oldendsubequations\endlinenomath}
  
\begin{document}

\title{Finite-size scaling analysis of the two-dimensional random transverse-field Ising ferromagnet}

\author{Jiwon Choi}
\affiliation{Department of Physics, Pukyong National University, Busan 48513, Korea}
\author{Seung Ki Baek}
 \email{seungki@pknu.ac.kr}
\affiliation{Department of Scientific Computing, Pukyong National University, Busan 48513, Korea}

\date{\today}

\begin{abstract}
The random transverse-field Ising ferromagnet (RTFIF) is a highly disordered quantum system which contains randomness in the coupling strengths as well as in the transverse-field strengths.
In one dimension, the critical properties are governed by an infinite-randomness fixed point (IRFP), and renormalization-group studies argue that the two-dimensional (2D) model is also governed by an IRFP. However, even the location of the critical point remains unsettled among quantum Monte Carlo (QMC) studies. In this work, we perform extensive QMC simulations to locate the quantum critical point and attempt a finite-size scaling analysis to observe the critical behavior.
We estimate the critical field strength of the 2D RTFIF as $\Gamma_c = 7.52(2)$, together with critical exponents such as $\beta=1.5(3)$, $\nu = 1.6(3)$, and $z=3.3(3)$ or $\psi=0.50(3)$.
We have also considered the McCoy-Wu model, which has randomness in the ferromagnetic coupling strengths but not in the transverse-field strength. Our QMC calculation shows that the critical behavior of the 2D McCoy-Wu model is closer to that of the 2D transverse-field Ising spin glass than to that of the 2D RTFIF. These numerical findings enhance our understanding of disordered 2D quantum systems.
\end{abstract}

\maketitle

\section{Introduction}

Novel phases of matter in disordered materials have provided fruitful insights for theoretical physics~\cite{mezard1987spin} as well as for information processing through neural networks~\cite{nishimori2001statistical,huang2021statistical}.
If we shift our attention to quantum phase transitions~\cite{sachdev1999quantum}, we may consider a natural generalization of the Ising model as follows:
\begin{equation}
    H = -\sum_{\left< i,j \right>} J_{ij} \hat{\sigma}_i^z \hat{\sigma}_j^z - \sum_i \Gamma_i \hat{\sigma}_i^x,
    \label{eq:hamiltonian}
\end{equation}
where $\hat{\sigma}_i^\alpha$ denotes the Pauli spin matrix in the $\alpha$ direction at site $i$. As usual, the first summation runs over the nearest neighbors. The coupling strength $J_{ij}$ between spin sites $i$ and $j$ is drawn from the distribution $P_\text{coupling}(J_{ij})$ to be specified below, and the transverse-field strength $\Gamma_i$ at site $i$ is drawn from another distribution, $P_\text{field}(\Gamma_i)$.
An important special case of Eq.~\eqref{eq:hamiltonian} is the celebrated McCoy-Wu model~\cite{mccoy1968theory,*mccoy1969theory}, which is equivalent to a one-dimensional (1D) disordered quantum spin chain as in Eq.~\eqref{eq:hamiltonian}, but with a uniform transverse-field strength, i.e., corresponding to $P_\text{field}(\Gamma_i) = \delta(\Gamma_i - \Gamma)$.
Although introduced as a theoretical extension of the classical Ising system to a quantum-mechanical one, Eq.~\eqref{eq:hamiltonian} now has an experimental realization in one dimension, where $P_\text{coupling}(J_{ij})$ and $P_\text{field}(\Gamma_i)$ are both Lorentzian~\cite{li2021spin}.

The 1D random transverse-field Ising ferromagnet (RTFIF) defined by Eq.~\eqref{eq:hamiltonian} has been successfully put into the renormalization-group (RG) context~\cite{fisher1992random,*fisher1995critical}. It has turned out that $P_\text{coupling}$ and $P_\text{field}$ renormalize to broader ones as the RG procedure continues, which means that the randomness in $J_{ij}$ and $\Gamma_i$ will eventually dominate the macroscopic behavior of the system. In other words, the system becomes more and more random as we move to larger and larger length scales, so the RG procedure eventually arrives at an infinite-randomness fixed point (IRFP)~\cite{fisher1999phase}.
A characteristic feature of such strong randomness is the difference between the \emph{average} and \emph{typical} correlations: Suppose that we look at the disordered phase in the vicinity of the quantum critical point. Although the correlations between almost all pairs of spins decay with a typical length scale $\xi_\text{typ}$, anomalously long spin clusters may form with small probabilities in the presence of strong disorder. As such rare events contribute to the \emph{average} correlations, the ``true'' correlation length $\xi$, which is the macroscopically measurable one, generally differs from $\xi_\text{typ}$ in the following way:
\begin{equation}
    \xi_\text{typ} \ll \xi^{1-\psi} \ll \xi,
\end{equation}
where $\psi$ quantifies the deviation from the non-random behavior.
Whereas the energy gap of a pure quantum spin system scales with the length scale $L$ as $\Omega \sim L^{-z}$, where $z$ is the dynamical critical exponent, the 1D RTFIF shows a striking difference because $\ln \Omega \sim -L^{\psi}$, which means that $z$ is effectively infinite.
We could alternatively say that the typical correlation between spins $i$ and $j$ still decay as $C_{ij} \sim \exp\left( -\left\vert \vec{r}_i - \vec{r}_j \right\vert / \xi_\text{typ} \right)$ even at the critical point, where $\vec{r}_k$ means the position of spin $k$. By contrast, their average behaves as $\overline{C_{ij}} \sim \left\vert \vec{r}_i - \vec{r}_j \right\vert^{-\eta}$.
Although this looks similar to the conventional scaling form, the statistics behind this average quantity is entirely different because of the existence of the rare events.

The same scenario has been believed to hold in higher dimensions: Most of the existing results for the two-dimensional (2D) RTFIF have been obtained either from the world-line quantum Monte Carlo (QMC) or from the strong-disorder RG (SDRG).
In those studies, the coupling strengths and transverse fields are drawn from 
\begin{subequations}
\begin{align}
    P_\text{coupling}(J_{ij}) &= \Theta(J_{ij}) \Theta(1-J_{ij}),\label{eq:jdist}\\
    P_\text{field}(\Gamma_i) &= \Gamma^{-1} \Theta(\Gamma_i) \Theta(\Gamma-\Gamma_i),
\end{align}
\label{eq:dist}
\end{subequations}
respectively, where $\Theta$ is the Heaviside step function and $\Gamma$ is the control parameter that defines the upper bound of $\Gamma_i$. The quantum phase transition occurs at $\Gamma = \Gamma_c$.
As summarized in Table~\ref{tab:earlier}, those studies consistently report $z=\infty$ and $\psi>0$ in agreement with the IRFP scenario.
The only exception is the cavity method~\cite{dimitrova2011cavity}, according to which $z$ is predicted to be finite at the critical point.
Table~\ref{tab:earlier} additionally shows numerical estimates of the critical exponent $\beta$, which describes the spontaneous magnetization as $M \propto \epsilon^\beta$, where $\epsilon$ is the distance from the critical point in the ordered phase.
It is worth mentioning that numerical results of the critical 2D random contact process~\cite{hooyberghs2003strong,vojta2009infinite} support the idea that it belongs to the 2D RTFIF universality class (Table~\ref{tab:universal}), although it is a non-equilibrium phenomenon. The random walk in a two-dimensional quenched random potential shows similar behavior~\cite{monthus2010random}, but its relation to the RTFIF is unclear.

\begin{table}[h]
\caption{\label{tab:earlier} Earlier results on the 2D RTFIF, together with our results in the last row. The last column has been calculated from $\beta = \nu \eta /2$~\cite{fisher1999phase}, and the results in the third last row have been obtained by using the projected cavity mapping~\cite{dimitrova2011cavity}. This table presents the data in chronological order and does not contain studies with different distributions for $J_{ij}$~\cite{karevski2001random,lin2007entanglement}. In the last row, we report $z$ and $\psi$ depending on which of the conventional and activated scaling scenarios is used. Note that we cannot have both at the same time because $z$ must be infinite if $\psi>0$.}
\begin{ruledtabular}
\begin{tabular}{ccccccc}
Method & $\Gamma_c$ & $z$ & $\psi$ & $\nu$ & $\eta$ & $\beta$\\\hline
QMC~\cite{pich1998critical} & 4.2(2) & $\infty$ & 0.42 & - &1.95 & -\\
SDRG~\cite{motrunich2000infinite} & - & $\infty$ & 0.42(6) & 1.07(15) & 2.0(2) & 1.1(2)\\
SDRG~\cite{lin2000numerical} & 5.3 & - & 0.5 & - & - \\
QMC~\cite{lin2002strongly} & 7.5 & $\infty$ & 0.58 & 1.33 & 2.3 & 1.53\\
SDRG~\cite{lin2002strongly} & 5.4(1) & $\infty$ & 0.48 & 1.25 & 2.06 & 1.29\\
SDRG~\cite{yu2008entanglement} & 5.35 & $\infty$ & - & - & 2.02(10) & -\\
SDRG~\cite{kovacs2009critical} & 5.344(26) & - & 0.51(3) & 1.25(8) & 1.99(3) & 1.24(8)\\
SDRG~\cite{kovacs2010renormalization} & 5.3570(5) & $\infty$ & 0.48(2) & 1.24(2)& 1.964(30)  & 1.22(4)\\
Cavity~\cite{dimitrova2011cavity} & 7.5 & 3.03 & - & - & -  & -\\
Boundary SDRG~\cite{monthus2012random} & 5.15 & - & 0.49 & 1.32 & - & -\\
This work & 7.52(2) & 3.3(3) & 0.50(3) & 1.6(3) & - & 1.5(3)
\end{tabular}
\end{ruledtabular}
\end{table}

\begin{table}[h]
\caption{\label{tab:universal}
Critical exponents from the 2D random contact process~\cite{vojta2009infinite}, which is believed to belong to the 2D RTFIF universality class. For comparison with Table~\ref{tab:earlier}, we have calculated $\eta$ by assuming the following scaling relation: $\eta = 2\beta/\nu_\perp$~\cite{fisher1999phase}.}
\begin{ruledtabular}
\begin{tabular}{ccccc}
 Method & $\psi$ & $\nu_\perp$ & $\eta$ & $\beta$\\
\hline
Monte Carlo & 0.51(6) & 1.20(15) & 2.0(5) & 1.15(15)\\
\end{tabular}
\end{ruledtabular}
\end{table}

Considering the fundamental importance of the RTFIF in the context of disordered quantum systems, we find it puzzling that little consensus exists about the critical point $\Gamma_c$ in two dimensions, even within the world-line QMC studies (Table~\ref{tab:earlier}).
Note that SDRG predicts $\Gamma_c \approx 5$ all the way through, which lies between the two QMC estimates: If $\Gamma_c = 4.2(2)$~\cite{pich1998critical}, SDRG could be said to underestimate the randomness in transverse fields but that it nevertheless leads to the IRFP. Or, if $\Gamma_c \approx 7.5$~\cite{lin2002strongly}, which coincides with the prediction of the projected cavity mapping~\cite{dimitrova2011cavity}, we could instead say that SDRG overestimates randomness. One might even argue that the 2D IRFP scenario is a consequence of such systematic overestimation.
For this reason, we wish to begin by locating the critical point through extensive QMC calculations for sampling rare events. We will then carry out finite-size scaling (FSS) at the critical point so as to estimate the critical exponents.

\section{Methods}

\subsection{World-line Monte Carlo method}

The world-line QMC method maps a $d$-dimensional quantum system to the corresponding $(d+1)$-dimensional classical system by adding an imaginary-time axis through the Suzuki-Trotter decomposition~\cite{gubernatis2016quantum}.
We have implemented the continuous imaginary-time code~\cite{kawashima2004recent} by modifying our previous one~\cite{baek2011quantum} to handle $L \times L$ square lattices with random couplings and random field strengths as given in Eq.~\eqref{eq:dist}. The periodic boundary conditions are imposed in both the spatial directions as well as in the imaginary-time direction.

In the continuous imaginary-time QMC method~\cite{rieger1999application}, each quantum spin $i$ is represented as a line along the imaginary-time axis, and the transverse field $\Gamma_i$ flips the spin direction from up to down or vice versa, cutting the line into segments of spin up and down. The positions of the cuts are given by a Poisson process with intensity $\Gamma_i$. Likewise, if two neighboring segments $i$ and $j$ are pointing in the same spin direction and have an overlap along the imaginary-time axis, we insert connections between them with a Poisson process with the intensity $2J_{ij}$ so that they can be flipped together by the Wolff-typed cluster update algorithm~\cite{blote2002cluster}.
The insertion of cuts and connections introduces a high degree of heterogeneity in the system:
Recall that if a line is broken into segments by a Poisson process with the intensity $\lambda$, the probability to find a segment longer than $t$ is $e^{-\lambda t}$, which implies that the segment length $\tau$ is roughly of an order of $\lambda^{-1}$. If $\lambda$ itself is a random variable drawn from a uniform distribution $p_\text{intensity}(\lambda)$ as in Eq.~\eqref{eq:dist}, the distribution $p_\text{length}(\tau)$ becomes extremely broad because it is given as
\begin{equation}
    p_\text{length}(\tau) = p_\text{intensity}(\lambda) \left\vert \frac{d\lambda}{d\tau} \right\vert \propto \tau^{-2},
\end{equation}
whose mean diverges~\cite{hidalgo2006conditions,*masuda2018gillespie}. Such large correlation in the imaginary-time direction is the origin of Griffiths singularities in quantum disordered systems~\cite{rieger1996griffiths}.

\subsection{Observables}

The first measurable quantity is total magnetization defined as
\begin{equation}
    m \equiv \frac1{L^2 b}\left[\left\langle\left\vert\sum_i\int_0^b\sigma_i^z(t)dt\right\vert\right\rangle\right],
    \label{eq:m}
\end{equation}
where $b \equiv (k_B T)^{-1}$ and $\sigma_i^z(t) \in \{+1, -1\}$ is the value of spin $i$ projected onto the $z$ direction at imaginary time $t$.
An average over disorder is denoted by $[\cdots]$, and a statistical-mechanical averages for a given sample is by $\langle\cdots\rangle$.
The conventional FSS form is assumed to be
\begin{equation}
    m = L^{-\beta/\nu}\mathcal M[(\Gamma-\Gamma_c)L^{1/\nu},bL^{-z}],
    \label{eq:m_fss}
\end{equation}
where $\mathcal{M}$ denotes the scaling function. In the \emph{activated} scaling, which is characteristic of the IRFP, a more appropriate scaling variable would be $(\ln b) L^{-\psi}$ instead of $bL^{-z}$.
To locate the critical point, we use Binder's cumulant defined as
\begin{equation}
    U = 1-\left[\frac{\langle m^4\rangle}{3\langle m^2\rangle^2}\right],
    \label{eq:u}
\end{equation}
where the disorder average is taken over the ratio of ensemble-averaged moments~\cite{hong2006anomalous}.
Again, the conventional FSS form is given by the following ansatz:
\begin{equation}
    U = \mathcal U[(\Gamma-\Gamma_c)L^{1/\nu},bL^{-z}],
\end{equation}
where $\mathcal{U}$ is the corresponding scaling function.

\begin{figure}
    \centering
    \includegraphics[width=0.48\textwidth]{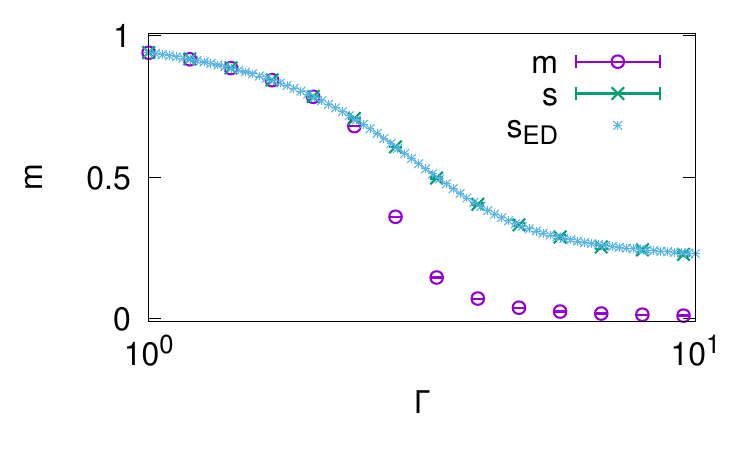}
    \caption{Comparison between QMC and ED for a single disorder realization with $L=4$. By $m$ and $s$, we mean Eqs.~\eqref{eq:m} and \eqref{eq:s}, respectively, whereas $s_{\text{ED}}$ denotes the spontaneous magnetization of the ground state obtained from ED. We set $b= 300$ in units of $k_B/J$ to make the contribution from the ground state dominant in the QMC calculation.}
    \label{fig:ed}
\end{figure}

We also measure the magnitude of spontaneous magnetization at a fixed position on the imaginary-time axis as follows~\cite{baek2011quantum}:
\begin{equation}
    s = \frac1{L^2}\left[\left\langle\left\vert\sum_i\sigma_i^z(0)\right\vert\right\rangle\right].
    \label{eq:s}
\end{equation}
Unlike Eq.~\eqref{eq:m}, it converges to a finite value as $b \to \infty$ at $\Gamma=\Gamma_c$. To see its meaning from a different angle,
let us consider the set of $z$-basis vectors, $\left\{ \left\vert s_l \right> \right\}$ with $l=1,\ldots, 2^N$, where $N\equiv L^2$ is the number of spins. If we represent the $j$th eigenstate as $\left\vert \Psi_j \right> \equiv \sum_l a_l^j \left\vert s_l \right>$ and denote its eigenvalue as $E_j$, Eq.~\eqref{eq:s} corresponds to $Z^{-1} \sum_j \left\vert s_j \right\vert e^{-bE_j}$,
where $Z$ is the partition function, and $\left\vert s_j \right\vert \equiv \sum_l \left\vert a_l^j \right\vert^2 \left\vert \left< s_l | \sum_i \sigma_i^z | s_l \right> \right\vert$ is the quantum-mechanical expectation value of $\sum_i \sigma_i^z$.
The eigenstates are obtainable from exact diagonalization (ED), and the results coincide with Eq.~\eqref{eq:s} (Fig.~\ref{fig:ed}). We expect that this quantity behaves similarly to $m$; therefore we write
\begin{equation}
    s = L^{-\beta/\nu}\mathcal S[(\Gamma-\Gamma_c)L^{1/\nu},bL^{-z}],
    \label{eq:s_fss}
\end{equation}
where the scaling function is denoted by $\mathcal{S}$.

\section{Results}

\subsection{Pure model}

\begin{figure}
    \centering
    \includegraphics[width=0.48\textwidth]{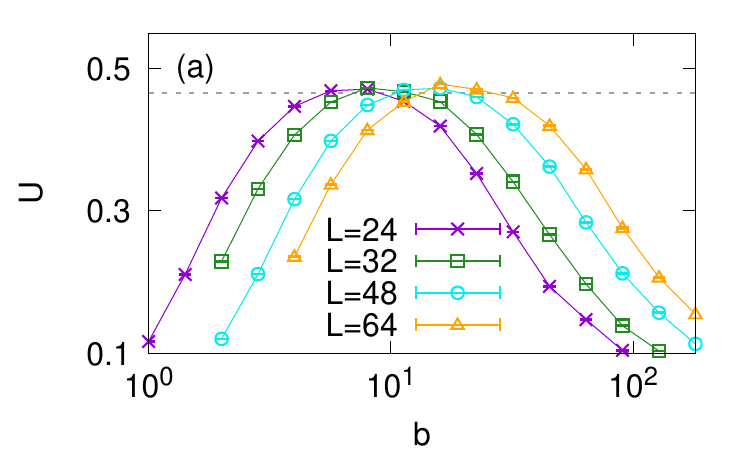}
    \includegraphics[width=0.48\textwidth]{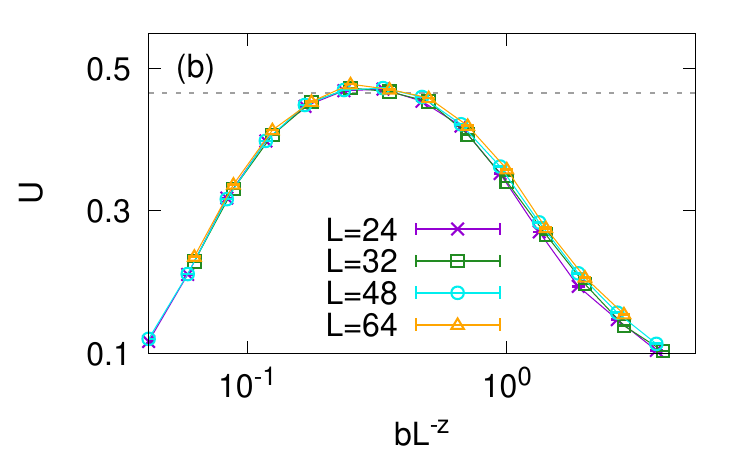}\\
    \includegraphics[width=0.48\textwidth]{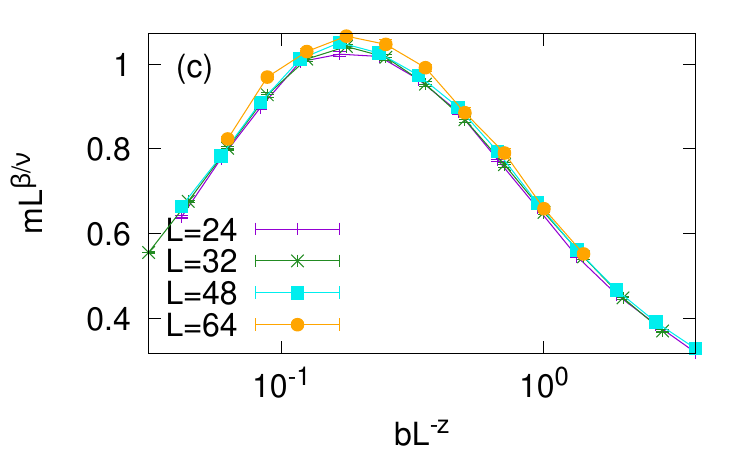}
    \includegraphics[width=0.48\textwidth]{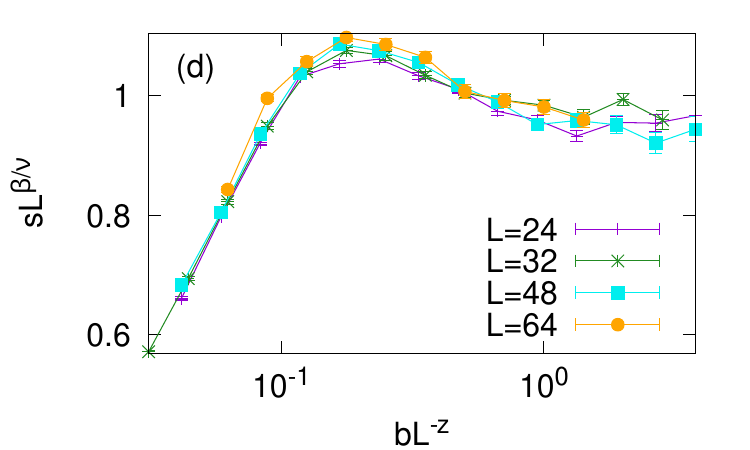}
    \caption{World-line QMC results for the pure 2D transverse-field Ising model [Eq.~\eqref{eq:pure}]. (a) Binder's cumulant $U$ [Eq.~\eqref{eq:u}] for various sizes at $\Gamma_c^\text{pure} = 3.044~330(6)$ in units of $J$~\cite{huang2020worm}. The horizontal line represents the universal amplitude of the 3D Ising model, $U^\ast = 0.465~48(5)$~\cite{ferrenberg2018pushing}. (b) Scaling collapse of $U$ with $z=1$. (c) Scaling collapse of total magnetization [Eq.~\eqref{eq:m}] with $\beta=0.3269(6)$ and $\nu=0.629~912(86)$ of the 3D Ising model~\cite{talapov1996magnetization,ferrenberg2018pushing}. (d) Scaling collapse of Eq.~\eqref{eq:s} with the same exponents.}
    \label{fig:pure}
\end{figure}

To test our code, we begin by simulating the pure 2D transverse-field Ising model, for which $P_\text{coupling}(J_{ij}) = \delta(J_{ij}-J)$ and $P_\text{field}(\Gamma_i) = \delta(\Gamma_i - \Gamma)$. The Hamiltonian is thus written as
\begin{equation}
    H_\text{pure} = -J \sum_{\left< i,j \right>} \hat{\sigma}_i^z \hat{\sigma}_j^z - \Gamma \sum_i \hat{\sigma}_i^x.
    \label{eq:pure}
\end{equation}
With $z=1$, the system is equivalent to the three-dimensional (3D) Ising model. Our QMC results are depicted in Fig.~\ref{fig:pure}, and they successfully reproduce the existing numerical estimates including the position of the critical point, $\Gamma_c^\text{pure}$.

\subsection{McCoy-Wu model}

By the McCoy-Wu model, we mean that the transverse fields in Eq.~\eqref{eq:hamiltonian} are fixed to a constant, i.e., $P_\text{field}(\Gamma_i) = \delta(\Gamma_i - \Gamma)$, whereas the disorder in coupling strengths is still given by Eq.~\eqref{eq:jdist}. The 2D version of the McCoy-Wu Hamiltonian is given as follows:
\begin{equation}
    H_\text{MW} = - \sum_{\left< i,j \right>} J_{ij} \hat{\sigma}_i^z \hat{\sigma}_j^z - \Gamma \sum_i \hat{\sigma}_i^x,
    \label{eq:mw}
\end{equation}
where the nearest neighbors are now defined on a square lattice.
Although no rigorous proof is available, this model has been believed to belong to the universality class of the 2D RTFIF in general: The scaling function of the 1D RTFIF obtained from RG is identical to the exact result of the 1D McCoy-Wu model~\cite{fisher1995critical}, and RG calculations support the conclusion that the 2D RTFIF and the 2D McCoy-Wu model exhibit the same critical behavior (Table~\ref{tab:mw}).

\begin{figure}
    \centering
    \includegraphics[width=0.48\textwidth]{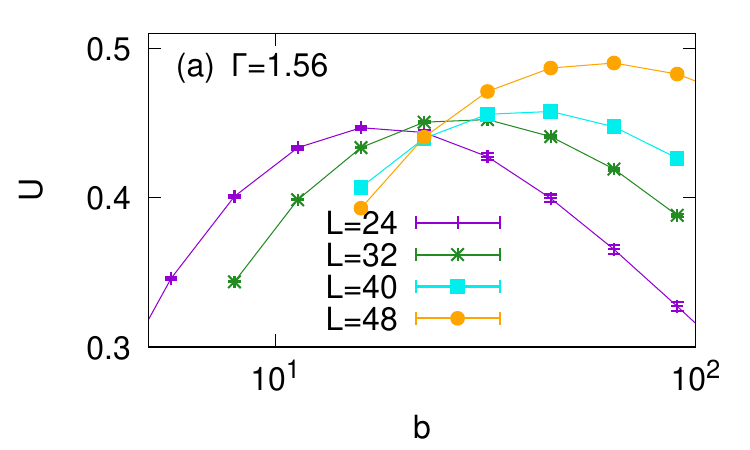}
    \includegraphics[width=0.48\textwidth]{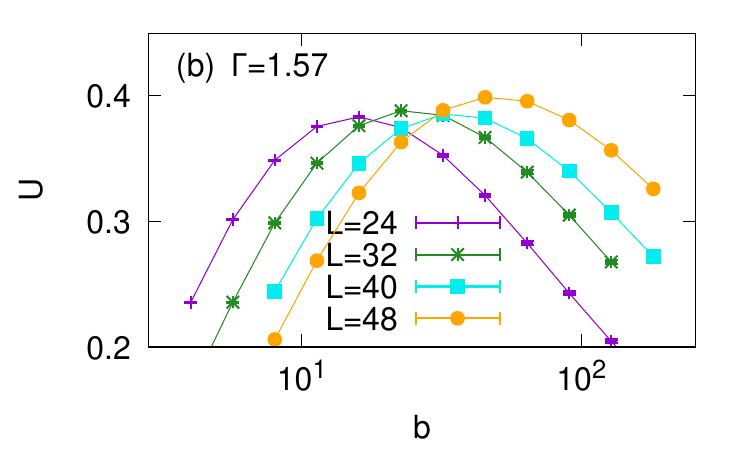}\\
    \includegraphics[width=0.48\textwidth]{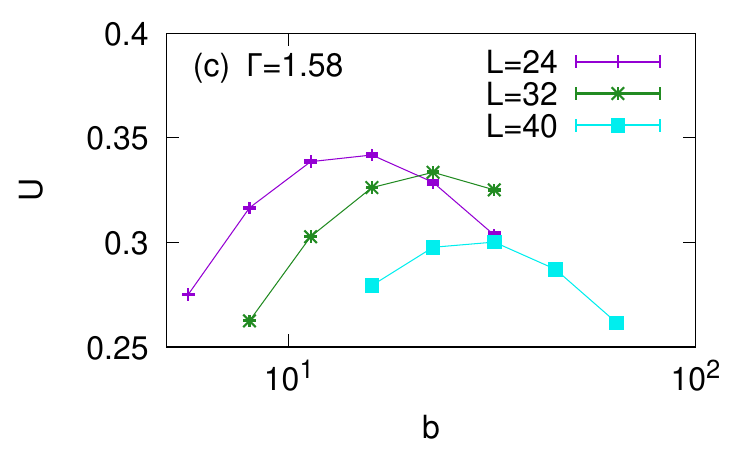}
    \caption{World-line QMC results for the 2D McCoy-Wu model [Eq.~\eqref{eq:mw}]. We depict Binder's cumulant at (a) $\Gamma=1.56$, (b) $\Gamma=1.57$, and (c) $\Gamma=1.58$ in units of $J$. These results locate the quantum critical point of the 2D McCoy-Wu model at $\Gamma=\Gamma_c^\text{MW} = 1.57(1)$.}
    \label{fig:mw}
\end{figure}

\begin{table}[h]
\caption{\label{tab:mw} Earlier results on the 2D McCoy-Wu model, together with our results in the last row. In the first row, we write the same value for $\psi$ as in Table~\ref{tab:earlier}, as well as for $\eta$, because these exponents have not been estimated separately from those of the RTFIF in Ref.~\onlinecite{kovacs2010renormalization}. To estimate $\beta$, we have used the following formula: $\beta = \nu \eta /2$~\cite{fisher1999phase}. The QMC study in the second last row has been conducted on a diluted Chimera graph whose degree of connectivity at each site is 3~\cite{nishimura2020griffiths}.}
\begin{ruledtabular}
\begin{tabular}{ccccccc}
Method & $\Gamma_c^\text{MW}$ & $z$ & $\psi$ & $\nu$ & $\eta$ & $\beta$\\\hline
SDRG~\cite{kovacs2010renormalization} & 0.84338(2) & $\infty$ & 0.48(2) & 1.23(2)& 1.964(30) & 1.21(3)\\
Real-space RG~\cite{miyazaki2013real} & 0.920 & $\infty$ & - & 1.20(6) & -  & -\\
QMC~\cite{nishimura2020griffiths} & 1.78 & $> 2.5(7)$ & - & 1.5 & -  & -\\
This work & 1.57(1) & 1.7(2) & 0.38(3) & 0.95(2) & - & 0.6(2)
\end{tabular}
\end{ruledtabular}
\end{table}

\begin{figure}
    \centering
    \includegraphics[width=0.48\textwidth]{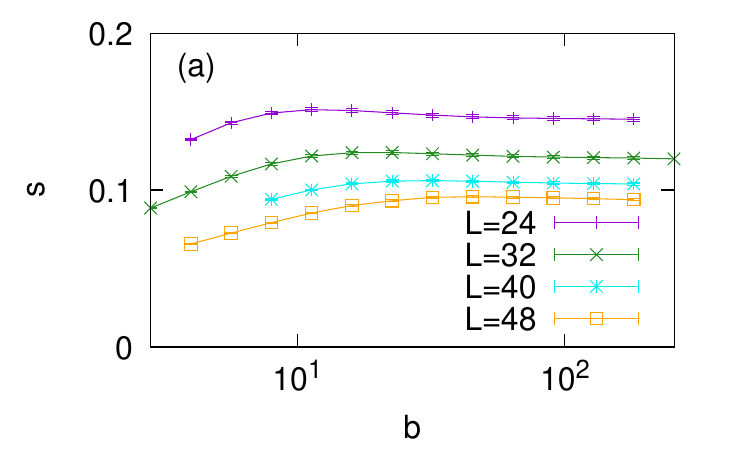}
    \includegraphics[width=0.48\textwidth]{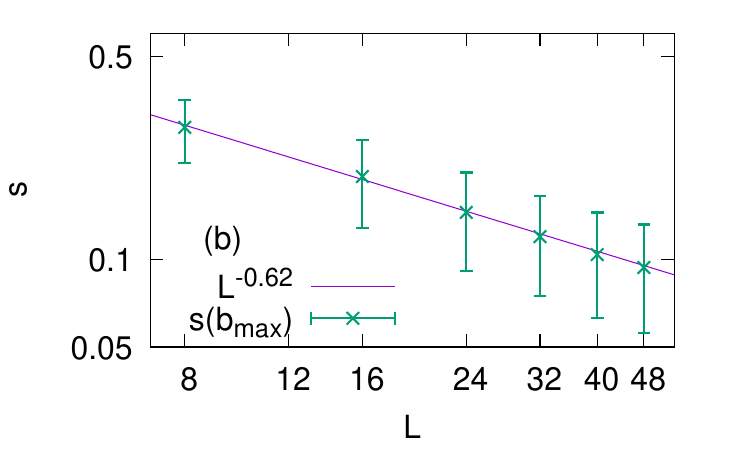}
    \caption{(a) Spontaneous magnetization of the 2D McCoy-Wu model at a fixed point on the imaginary-time axis. Note that this quantity reaches a plateau as $b$ grows. (b) The same observable at the largest available $b\sim2\times10^2$ for each system size. We have plotted $L^{-0.62}$ for comparison.}
    \label{fig:mw_smax}
\end{figure}

\begin{figure}
    \centering
    \includegraphics[width=0.48\textwidth]{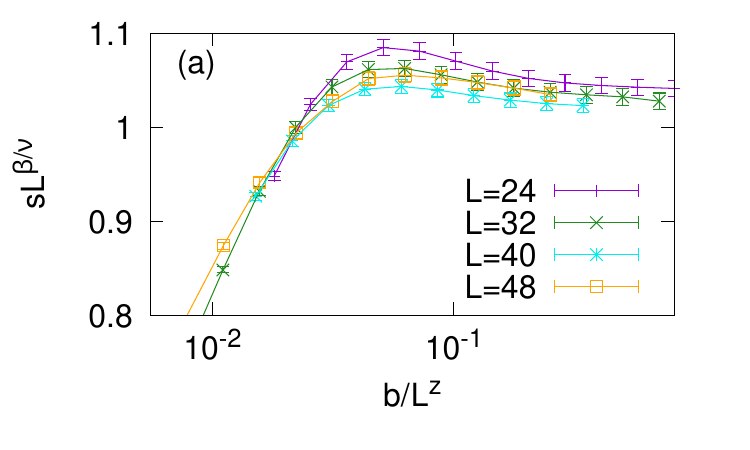}
    \includegraphics[width=0.48\textwidth]{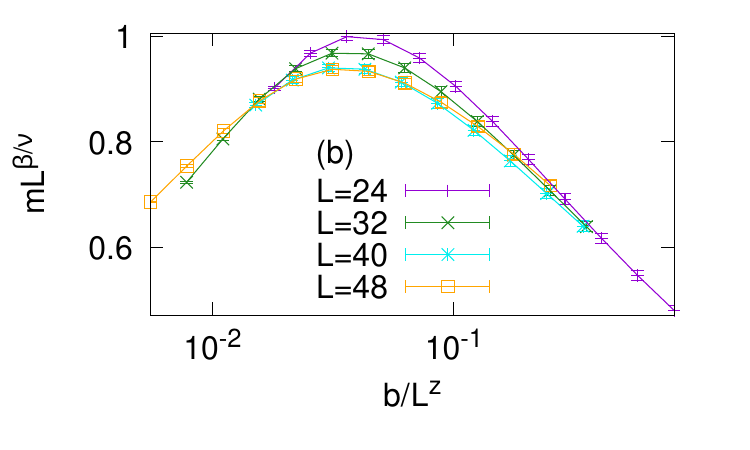}
    \includegraphics[width=0.48\textwidth]{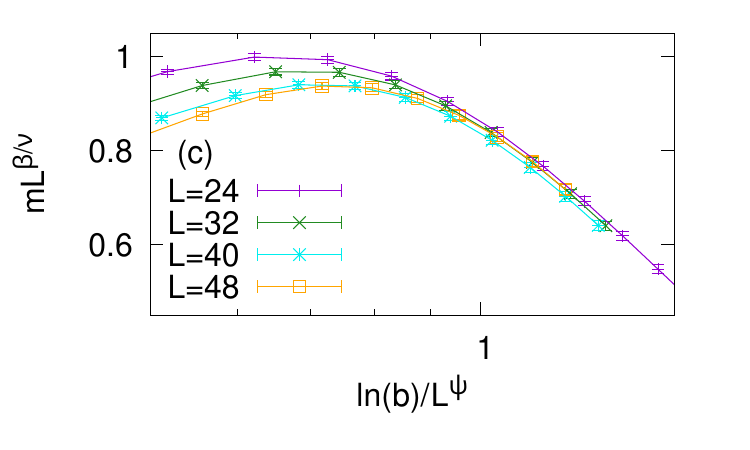}
    \includegraphics[width=0.48\textwidth]{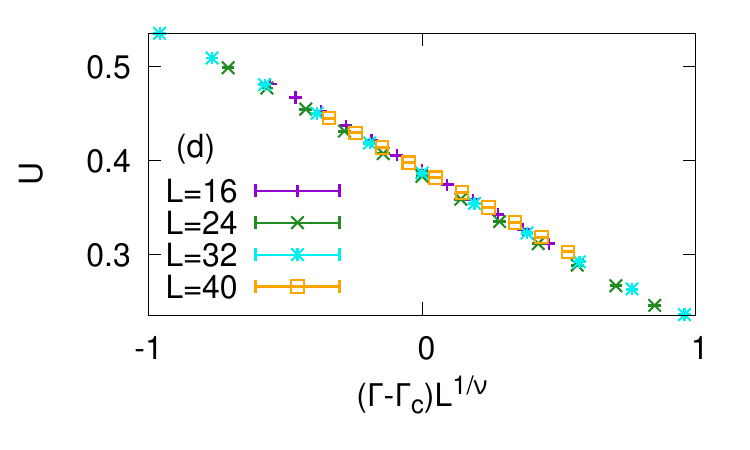}
    \caption{Scaling collapse of the 2D McCoy-Wu model at $\Gamma_c^{\text{MW}} = 1.57$. (a) By observing the large-$b$ behavior of $s$, we estimate $\beta/\nu = 0.62(20)$ [see also Fig.~\ref{fig:mw_smax}(b)]. (b) If we use this value, the large-$b$ behavior of $m$ is best described by $z=1.7(2)$ if we assume the conventional scaling or (c) $\psi=0.38(3)$ in the activated scaling form. (d) By choosing $b=b^\ast$ at which $U$ reaches a maximum for each given $L$, we increase the system size in both space and time.
    From the resulting FSS of Binder's cumulant $U$, we estimate $\nu = 0.95(2)$.}
    \label{fig:mw_fss}
\end{figure}

In Table~\ref{tab:mw}, we have listed existing results of the 2D McCoy-Wu model.
The QMC results in the second last row~\cite{nishimura2020griffiths} are likely to contain strong finite-size effects for the following reasons: The coupling strengths are uniformly drawn from $\{0, 0.2, 0.4, 0.6, 0.8, 1.0\}$ instead of the continuous unit interval, and the system sizes are limited to $8 \times L\times L\times M$, with $L \le 12$ and $M=150$, where $M$ is the number of time slices in the imaginary-time direction, and the factor of 8 is due to the topology of the diluted Chimera graph used in the quasi-2D simulation~\cite{nishimura2020griffiths}. It is not easy to compare the different structures with different sizes directly. On the one hand, the diluted Chimera graph has a smaller degree of connectivity than the 2D square lattice, which suggests that the ferromagnetic order will be more fragile. On the other hand, the periodic boundary conditions tends to underestimate fluctuations in such small systems~\cite{cardy1996scaling}.
In any case, the striking difference in $\Gamma_c^\text{MW}$ between the RG and QMC calculations suggests that QMC would give a higher value for the critical field strength than the SDRG calculation, which in turn lends some support to $\Gamma_c \approx 7.5$ in the 2D RTFIF case. By using the continuous imaginary-time QMC code, we estimate the critical point of the 2D McCoy-Wu model as $\Gamma_c^\text{MW} = 1.57(1)$ (Fig.~\ref{fig:mw}). For the 2D McCoy-Wu model, we have used more than $3\times 10^3$ disorder realizations for each system size.

At this critical point, we begin our FSS analysis by observing the behavior of $s$ (Fig.~\ref{fig:mw_smax}). In Eq.~\eqref{eq:s_fss}, this quantity is assumed to be a function of $b$ and $L$ at $\Gamma = \Gamma_c$, but it becomes insensitive to $b$ for sufficiently large values of $b$ [Fig.~\ref{fig:mw_smax}(a)]. We thus end up with a function of $L$ only. By fitting $s(L)$ to $L^{-\beta/\nu}$, we find $\beta/\nu = 0.62(20)$ [Fig.~\ref{fig:mw_smax}(b)]. As shown in Fig.~\ref{fig:pure}(c), the dynamic critical exponent $z$ (or $\psi$) can then be checked from the scaling collapse of $m L^{\beta/\nu}$ in the low-temperature region, based on the above estimate of $\beta/\nu$. Finally, to obtain $\nu$ separately, we measure Binder's cumulant by lowering the temperature according to $b \propto L^z$ (or $b \propto e^{L^\psi}$) as $L$ grows (Fig.~\ref{fig:mw_fss}).

A notable point in our numerical results is that the critical amplitude of $U$ is around $0.4$ [see Fig.~\ref{fig:mw}(b)], whereas that of the 2D RTFIF has been reported as close to $1/3$~\cite{pich1998critical}. Indeed, our FSS analysis at $\Gamma_c^\text{MW}$ gives $\beta=0.6(2)$, $\nu = 0.95(2)$, and $z=1.7(2)$ or $\psi=0.38(3)$ (Fig.~\ref{fig:mw_fss}), and these values clearly differ from the earlier estimates in the 2D RTFIF (Table~\ref{tab:earlier}). Interestingly, they are actually closer to those of the 2D transverse-field Ising spin glass (Table~\ref{tab:ig}). It implies that the existence of frustration could be irrelevant to the critical behavior of these 2D systems. Such irrelevance of frustration is obvious in a 1D chain because the coupling strengths and the transverse-field strengths can always be made positive by simple gauge transformation as long as no external field exists in the $z$ direction~\cite{fisher1995critical}, but it is far from trivial in 2D systems. Of course, this observed difference between the 2D RTFIF and the 2D McCoy-Wu model needs to be taken with care because we have dealt with small sizes whereas the true scaling behavior will manifest itself in the infinite-size limit. Even in the 1D McCoy-Wu model for which the IRFP behavior is well established, crossover effects can be dominant in finite-sized systems~\cite{crisanti1994random}.

Before proceeding, we note a controversy about the dynamic scaling behavior of the 2D transverse-field Ising spin glass: A recent study argues that it is governed by an IRFP~\cite{matoz2016unconventional} in contrast with an earlier numerical study~\cite{rieger1994zero}.
If the 2D McCoy-Wu model belongs to the universality class of the spin glass as suggested here, it may provide another window into this problem between the conventional and activated scaling forms because the McCoy-Wu model is a ferromagnetic system with no frustration, which is expected to reach equilibrium faster than the spin glass.
The validity of this conjectured universality, as well as a detailed comparison between the conventional and activated scaling forms, is left as a future work.

\begin{table}[h]
\caption{\label{tab:ig} Earlier world-line QMC results on the 2D transverse-field Ising spin glass. The Hamiltonian takes the same form as in Eq.~\eqref{eq:mw}, but the distribution of coupling strengths differs. By "Bimodal", we mean $P(J_{ij}) = \frac{1}{2}[\delta(J_{ij}-1) + \delta(J_{ij}+1)]$, and $\mathcal{N}(\mu, \sigma^2)$ denotes the Gaussian distribution with mean $\mu$ and variance $\sigma^2$. The QMC studies in this table are parametrized with $K \approx (1/2) \ln \coth (\Delta \tau \Gamma)$, which becomes exact in the limit of infinite imaginary-time slices, i.e., $\Delta \tau \to 0$, but all these studies have estimated the critical point $K_c$ with setting $\Delta \tau=1$~\cite{rieger1994zero,matoz2016unconventional}.}
\begin{ruledtabular}
\begin{tabular}{ccccccc}
$P_\text{coupling}(J_{ij})$ & Method & $\Gamma_c$ & $z$ & $\psi$ & $\nu$ \\\hline
$\mathcal{N}(0,1)$~\cite{rieger1994zero} & QMC & 0.608(3) & 1.50(5) & - & 1.0(1)\\
$\mathcal{N}(0,1)$~\cite{matoz2016unconventional} & QMC & 0.615(4) & 1.55 & 0.44(3) & 1.13(5)\\
Bimodal~\cite{matoz2016unconventional} & QMC & 0.638(1) & 1.7 & 0.46(1) & 1.2(4)\\
$\mathcal{N}(0,1)$~\cite{miyazaki2013real} & Real-space RG & 1.195 & - & - & 1.21(9)

\end{tabular}
\end{ruledtabular}
\end{table}

\subsection{Random transverse-field model}

\begin{figure}
    \centering
    \includegraphics[width=0.48\textwidth]{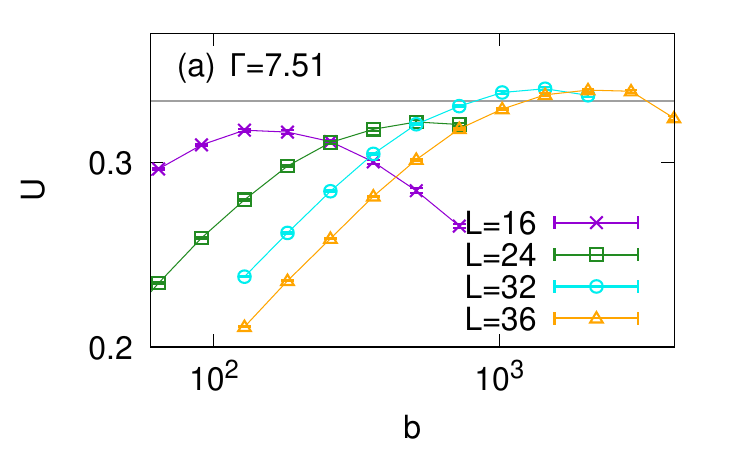}
    \includegraphics[width=0.48\textwidth]{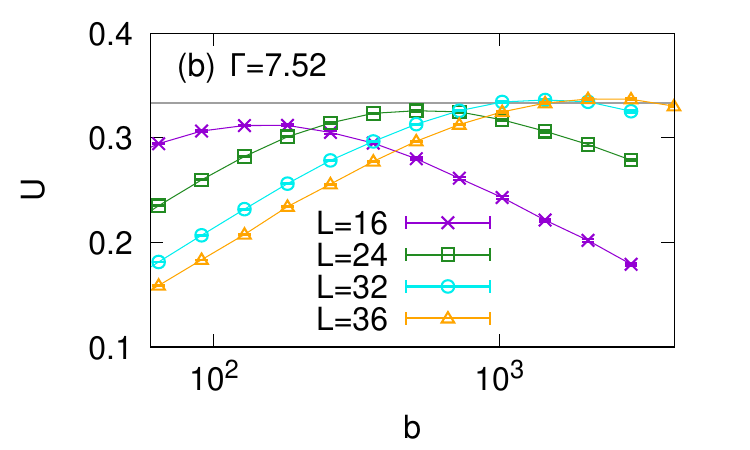}
    \includegraphics[width=0.48\textwidth]{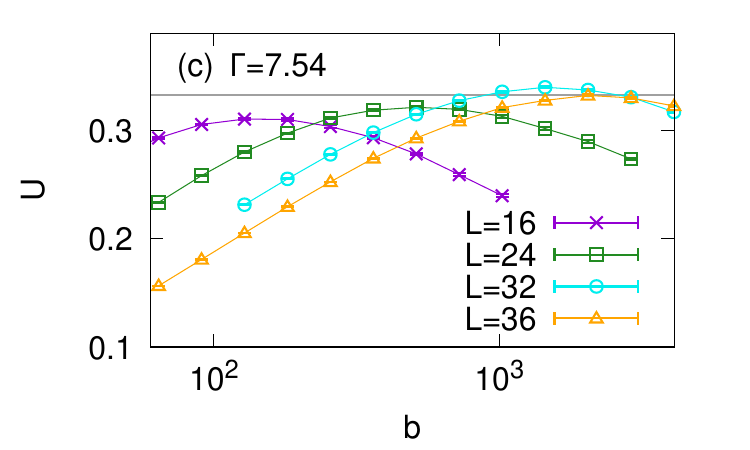}
    \caption{Binder's cumulant of 2D RTFIF at (a) $\Gamma = 7.51$, (b) $\Gamma = 7.52$, and (c) $\Gamma=7.54$ in units of $J$. The horizontal line represents the critical value of Binder's cumulant $U^\ast\approx0.33$ from Ref.~\onlinecite{pich1998critical}. This result locates the quantum critical point of the 2D RTFIF at $\Gamma_c=7.52$.}
    \label{fig:rtfif}
\end{figure}

Now we deal with the 2D RTFIF as defined by Eq.~\eqref{eq:hamiltonian}, where disorder exists in both coupling and the transverse field [Eq.~\eqref{eq:dist}].
The 2D RTFIF exhibits serious finite-size effects:
Figure~\ref{fig:rtfif} shows the world-line QMC results for Binder's cumulant at various values of $\Gamma$. For the 2D RTFIF, we have used more than $2\times 10^3$ disorder realizations for each system size. We can locate the critical point $\Gamma_c$ only by comparing the two largest sizes, $L=32$ and $36$. Our estimate is $\Gamma_c=7.52(2)$, which is consistent with $\Gamma_c=7.5$ in Ref.~\onlinecite{lin2002strongly}.

\begin{figure}
    \centering
    \includegraphics[width=0.48\textwidth]{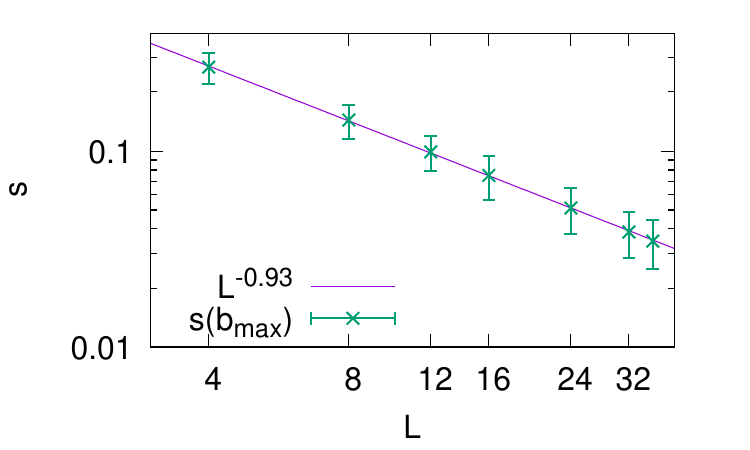}
    \caption{Spontaneous magnetization at a fixed point on the imaginary-time axis, when we take the largest available $b\sim4\times10^3$ for each system size. We have plotted $L^{-0.93}$ for comparison.}
    \label{fig:rtfif_betanu}
\end{figure}

We then estimate critical exponents by using FSS.
The spontaneous magnetization at a fixed imaginary-time slice [Eq.~\eqref{eq:s}] is a convenient observable in estimating $\beta/\nu$ because it converges to a finite value for sufficiently large $b$. For every $L$, our data points at the largest values of $b$ overlap within the error bars. By observing the behavior of $s$ as a function of $L$, we estimate $\beta/\nu=0.93(11)$ (Fig.~\ref{fig:rtfif_betanu}). Even if we observe only the two largest system sizes, it does not alter this result in any significant way.

\begin{figure}
    \centering
    \includegraphics[width=0.48\textwidth]{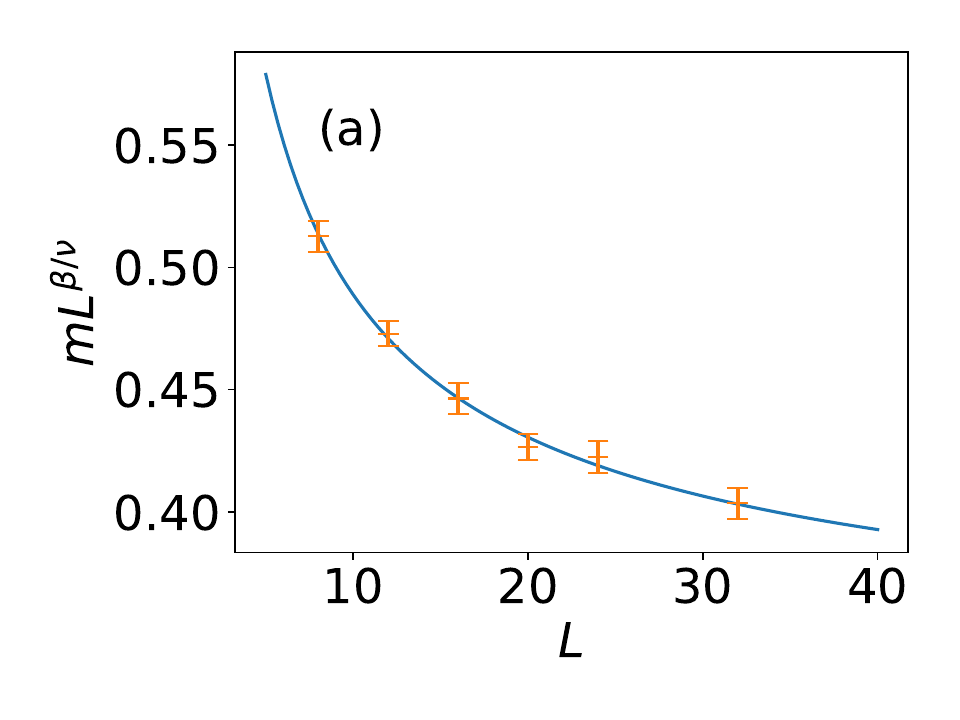}
    \includegraphics[width=0.48\textwidth]{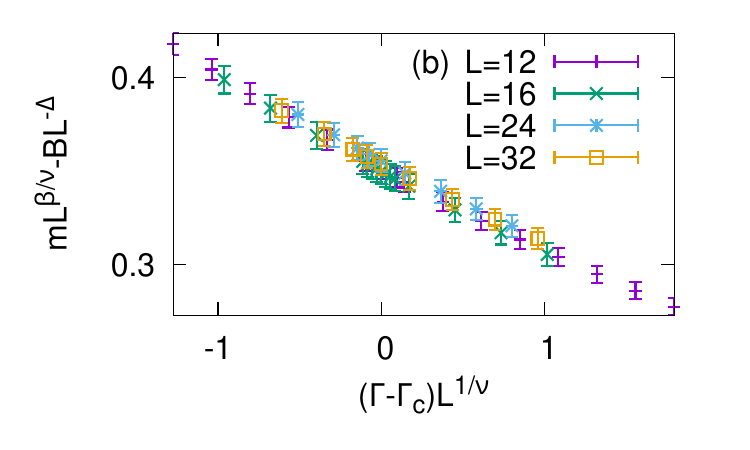}
    \caption{(a) Estimation of $A$, $B$, and $\Delta$ based on Eq.~\eqref{eq:correction}. The solid line depicts the fitting result with $A=0.32(6)$, $B=0.70(25)$, and $\Delta=0.63(33)$. (b) Estimation of the correlation-length exponent $\nu=1.6(3)$ when the sub-leading correction is included.}
    \label{fig:rtfif_nu}
\end{figure}

\begin{figure}
    \centering
    \includegraphics[width=0.48\textwidth]{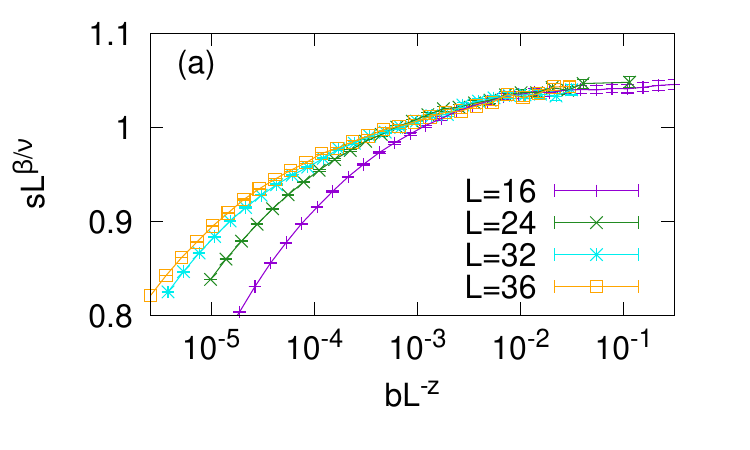}
    \includegraphics[width=0.48\textwidth]{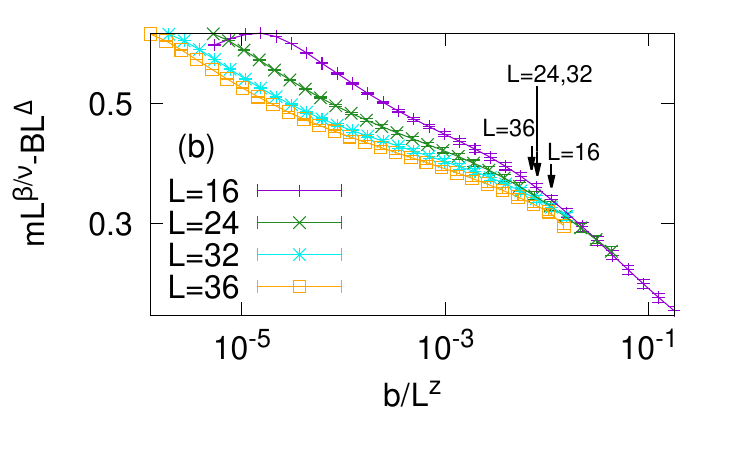}
    \includegraphics[width=0.48\textwidth]{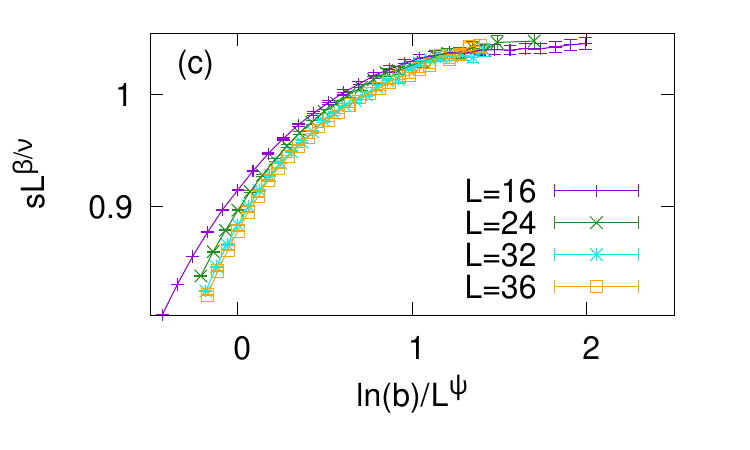}
    \includegraphics[width=0.48\textwidth]{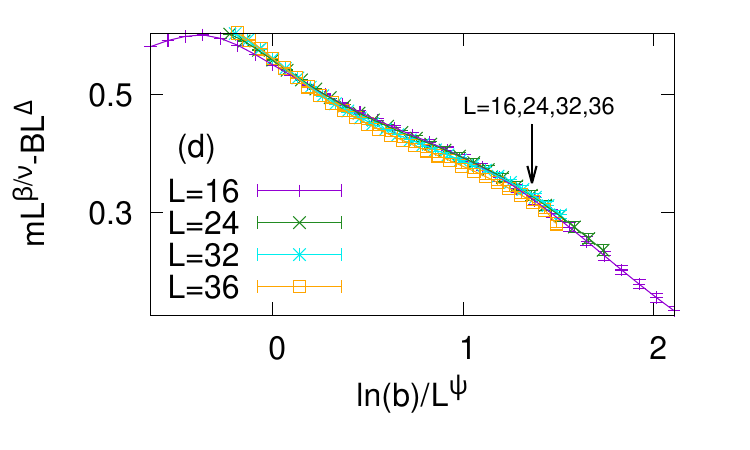}
    \caption{Scaling collapse of the 2D RTFIF at $\Gamma_c=7.52$. In the same way as in Fig.~\ref{fig:mw_fss}, (a) we estimate $\beta/\nu=0.93(11)$ by observing the large-$b$ behavior of $L=32$ and $36$. Based on this value, (b) we estimate $z=3.3(3)$ using the conventional scaling. If we assume the activated scaling, we estimate $\psi = 0.50(3)$ in panels (c) and (d). The vertical arrows indicate $b^\ast(L)$ at which Binder's cumulant reaches a maximum for each given $L$ (see Fig.~\ref{fig:rtfif}).}
    \label{fig:rtfif_fss}
\end{figure}

The total magnetization $m$ suffers more from finite-size effects. Let us assume the following sub-leading correction to the scaling:
\begin{equation}
    mL^{\beta/\nu} = A+BL^{-\Delta},
    \label{eq:correction}
\end{equation}
where $A$, $B$, and $\Delta$ are unknowns to be found by data fitting (Fig.~\ref{fig:rtfif_nu}). For sufficiently large values of $b$, we expect that the unknowns will converge to constants describing the zero-temperature behavior. Of our primary interest is $A$ which describes the leading contribution to scaling.
Let us use $\beta/\nu = 0.93$ to describe the large-$b$ behavior of $m$, with the sub-leading correction taken into account. If we choose the conventional scaling ansatz [Eq.~\eqref{eq:m_fss}], we have few data points in the scaling region, i.e., $bL^{-z} \gtrsim O(10^{-2})$, but $z=3.3(3)$ cannot be ruled out [see Fig.~\ref{fig:rtfif_fss}(b)]. We may compare this result with the prediction of the cavity method (Table.~\ref{tab:earlier}). If we alternatively assume the activated scaling, we get $\psi=0.50(3)$ [Fig.~\ref{fig:rtfif_fss}(d)]. The scaling collapse is more convincing in the activated scaling form, but we should check larger system sizes to reach a conclusion.

To estimate $\beta$ and $\nu$ separately, we perform FSS in space-time by increasing $L$ and lowering the temperature accordingly. Although our QMC results do not provide a clear-cut answer between the conventional and activated scaling hypotheses, a phenomenological procedure is still possible:
For each $L$ in Fig.~\ref{fig:rtfif}(b), let us choose $b^\ast(L)$ at which Binder's cumulant reaches a maximum. We lower the temperature according to $b^\ast(L)$ as $L$ increases.
Also by taking into account the sub-leading correction in Eq.~\eqref{eq:correction}, we estimate $\nu=1.6(3)$ (Fig.~\ref{fig:rtfif_nu}), which is consistent with an earlier QMC result~\cite{lin2002strongly}.

\begin{figure}
    \centering
    \includegraphics[width=0.48\textwidth]{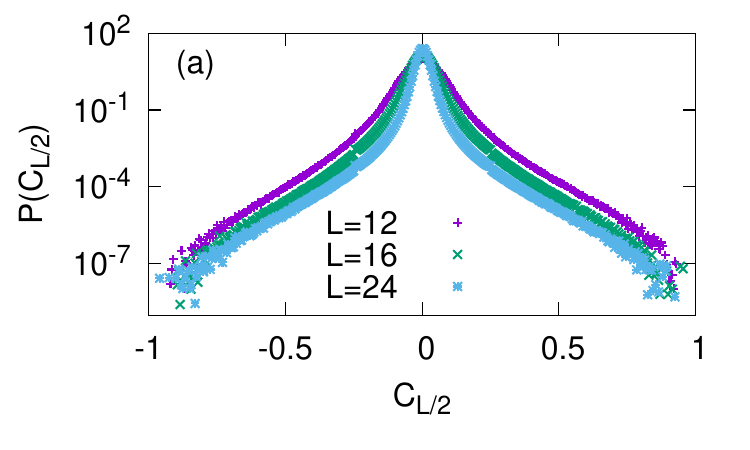}
    \includegraphics[width=0.48\textwidth]{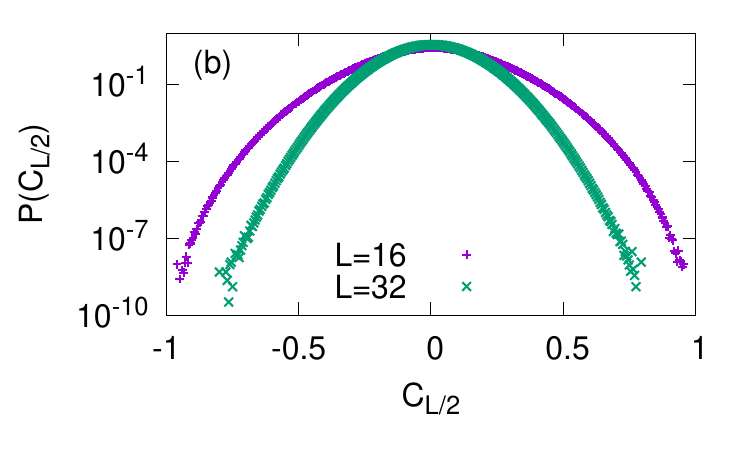}
    \caption{ Normalized histograms of $C_{L/2}$, the correlation between two spins at a distance of $L/2$. (a) For the 2D RTFIF at $b^\ast(L)$, the histogram develops long tails, signaling the existence of anomalously long spin clusters. (b) By contrast, the histogram takes the simple Gaussian form in the critical 2D McCoy-Wu model.
    }
    \label{fig:rtfif_corr}
\end{figure}

Between the conventional and activated scaling scenarios, however, our results from the FSS are not enough to rule out any of them, although it is easier to fit the data in the activated scaling form. Therefore, we have checked the two-spin correlation at a large distance to detect the presence of anomalously long spin clusters. The correlation between spin $i$ and $j$ is defined as~\cite{lin2002strongly}
\begin{equation}
    C(i,j) = \frac1b\int_0^b\sigma_i^z(t)\sigma_j^z(t)dt - \frac1{b^2}\int_0^b\sigma_i^z(t)dt\cdot\int_0^b\sigma_j^z(t')dt'.
\end{equation}
We average this quantity over every pair of $i$ and $j$ at a distance of $L/2$, separated either vertically or horizontally in the square lattice, to obtain $C_{L/2}$.
In Fig.~\ref{fig:rtfif_corr}(a), we draw normalized histograms of $C_{L/2}$ in the 2D RTFIF at the critical point, obtained from $10^3$ disorder realizations for $L=12$ and from $2\times 10^3$ disorder realizations for $L\ge 16$. We have also repeated the measurement of $C_{L/2}$ over $5\times 10^5$ Monte Carlo steps to improve the statistics for each disorder realization. The resulting histograms have long tails, indicating the emergence of anomalously long spin clusters, and this pattern survives in large systems. The tail parts will thus make the average behavior different from the typical one, which is a characteristic feature of the IRFP. Note that this histogram is sharply contrasted with that in the McCoy-Wu model, where such long tails are totally absent [Fig.~\ref{fig:rtfif_corr}(b)]. This comparison supports our earlier observation that these two models do not belong to the same universality class.

\section{Discussion and Summary}

To summarize, we have carried out FSS analysis to understand the critical properties of the 2D RTFIF by conducting the world-line QMC simulations. The first contribution of this work is that we have found the critical field strength $\Gamma_c = 7.52(2)$, confirming an earlier, almost forgotten, result~\cite{lin2002strongly}.
Regarding this estimate of the critical point, the cavity method~\cite{dimitrova2011cavity} shows striking agreement, which is non-trivial because the cavity method assumes a locally tree-like structure, whereas we are dealing with a 2D system. The geometric aspects of the critical RTFIF would thus be worth investigating, such as whether the performance of the cavity method is related to the percolation threshold of the underlying lattice structure as well as to the distributions of disorder [Eq.~\eqref{eq:dist}].
In contrast with the cavity method, the SDRG tends to overestimate disorder (see Table~\ref{tab:earlier}).

The earlier QMC calculation~\cite{lin2002strongly} and the cavity method~\cite{dimitrova2011cavity}, however, yield different predictions on the dynamic scaling behavior: The former supports the activated scaling with $\psi >0$, whereas the latter does the conventional scaling with $z \approx 3$. Overall, our results favor the activated scaling scenario in the sense that the system exhibits a characteristic phenomenon of the IRFP when we look at the spin-spin correlation [see Fig.~\ref{fig:rtfif_corr}(a)].

As for other critical exponents, we have estimated $\beta = 1.5(3)$ and $\nu = 1.6(3)$, both of which are consistent with the existing ones~\cite{lin2002strongly}.
The main difficulty in FSS arises from the strong finite-size effects: Even for $L=36$, the inverse temperature needs to be as large as $O(10^3)$ to access the critical region (Fig.~\ref{fig:rtfif}), and it will increase more drastically for larger system sizes, whether the peak position follows $b^\ast \sim L^z$ according to the conventional scaling or $b^\ast \sim \exp(L^{\psi})$ according to the activated one. The high value of the critical field strength $\Gamma_c \sim O(10)$ causes a heavy memory load because a large number of cuts must be introduced to such long world lines.
We have added the subleading correction to cope with the finite-size effects, but numerical approaches other than the world-line method would also be desirable to probe the low-temperature behavior of larger systems.

In addition, according to our results, the 2D McCoy-Wu model belongs to a different universality class from that of the 2D RTFIF, differently from the 1D case.
Although the SDRG makes the same prediction for both the models, it is not supported by our QMC calculation: The distribution of the transverse-field strength seems to make a qualitative difference in 2D systems. Furthermore, we have suggested that the 2D McCoy-Wu model could rather belong to the universality class of the 2D transverse-field Ising spin glass. Whether they are governed by the same fixed point regardless of the existence of frustration is an intriguing question to be scrutinized in detail.

From a broader perspective, our work shows that we still need more precise understanding of 2D quantum systems in the presence of disorder. The picture of Anderson localization implies that disorder will dominate the physics of 2D quantum systems at large. However, as we have seen by comparing the 2D McCoy-Wu model and the 2D RTFIF, a variety of different aspects may exist in this picture, and the lack of rigorous theoretical answers calls for extensive numerical experiments to understand the interplay between disorder and quantumness.

\begin{acknowledgments}
We gratefully acknowledge valuable comments from Heiko Rieger.
This work was supported by a research grant from Pukyong National
University (2022).
\end{acknowledgments}


\bibliographystyle{apsrev4-2}
\bibliography{apssamp}

\end{document}